\pdfoutput=1

\documentclass[pra,twocolumn,superscriptaddress,letterpaper]{revtex4}

\usepackage{graphicx,float}
\usepackage{epsfig}
\usepackage{dcolumn}
\usepackage{bm,mathptmx}
\usepackage{amsmath}
\usepackage{amssymb}
\usepackage[colorlinks=true,citecolor=blue,urlcolor=blue]{hyperref}

\newcommand{\ket}[1]{\ensuremath{|\,{#1}\,\rangle}}
\newcommand{\bra}[1]{\ensuremath{\langle\,{#1}\,|}}

\begin{document}

%%%%%%%%%%%%%%%%%%%%%%%%%%%%%%%%%%%%%%%%%%%%%%%%%%%%%%%%%%%%%%%%%

\title{Experimental implementation of an eight-dimensional Kochen-Specker set and observation of its connection with the Greenberger-Horne-Zeilinger theorem}

%%%%%%%%%%%%%%%%%%%%%%%%%%%%%%%%%%%%%%%%%%%%%%%%%%%%%%%%%%%%%%%%%

\author{Gustavo~Ca\~nas}
\author{Sebasti\'an~Etcheverry}
\author{Esteban~S.~G\'omez}
\affiliation{Departamento de F\'isica, Universidad de Concepci\'on, 160-C Concepci\'on, Chile}
\affiliation{Center for Optics and Photonics, Universidad de Concepci\'on, Concepci\'on, Chile}
\affiliation{MSI-Nucleus for Advanced Optics, Universidad de Concepci\'on, Concepci\'on, Chile}

\author{Carlos~Saavedra}
\affiliation{Departamento de F\'isica, Universidad de Concepci\'on, 160-C Concepci\'on, Chile}
\affiliation{Center for Optics and Photonics, Universidad de Concepci\'on, Concepci\'on, Chile}

\author{Guilherme~B.~Xavier}
\affiliation{Center for Optics and Photonics, Universidad de Concepci\'on, Concepci\'on, Chile}
\affiliation{MSI-Nucleus for Advanced Optics, Universidad de Concepci\'on, Concepci\'on, Chile}
\affiliation{Departamento de Ingenier\'ia El\'ectrica, Universidad de Concepci\'on, 160-C Concepci\'on, Chile}

\author{Gustavo~Lima}
\email{glima@udec.cl}
\affiliation{Departamento de F\'isica, Universidad de Concepci\'on, 160-C Concepci\'on, Chile}
\affiliation{Center for Optics and Photonics, Universidad de Concepci\'on, Concepci\'on, Chile}
\affiliation{MSI-Nucleus for Advanced Optics, Universidad de Concepci\'on, Concepci\'on, Chile}

\author{Ad\'an~Cabello}
\email{adan@us.es}
\affiliation{Departamento de F\'isica Aplicada II, Universidad de Sevilla, E-41012, Sevilla, Spain.}

%%%%%%%%%%%%%%%%%%%%%%%%%%%%%%%%%%%%%%%%%%%%%%%%%%%%%%%%%%%%%%%%%%%

\begin{abstract}
 For eight-dimensional quantum systems there is a Kochen-Specker (KS) set of 40~quantum yes-no tests that is related to the Greenberger-Horne-Zeilinger (GHZ) proof of Bell's theorem. Here we experimentally implement this KS set using an eight-dimensional Hilbert space spanned by the transverse momentum of single photons. We show that the experimental results of these tests violate a state-independent noncontextuality inequality. In addition, we show that, if the system is prepared in states that are formally equivalent to a three-qubit GHZ and $W$ states, then the results of a subset of 16~tests violate a noncontextuality inequality that is formally equivalent to the three-party Mermin's Bell inequality, but for single eight-dimensional quantum systems. These experimental results highlight the connection between quantum contextuality and nonlocality for eight-dimensional quantum systems.
\end{abstract}

%%%%%%%%%%%%%%%%%%%%%%%%%%%%%%%%%%%%%%%%%%%%%%%%%%%%%%%%%%%%%%%%%%%

% This version: May 16, 2014 (Sevilla).

\date{\today}

\maketitle

%%%%%%%%%%%%%%%%%%%%%%%%%%%%%%%%%%%%%%%%%%%%%%%%%%%%%%%%%%%%%%%%%%%

\section{Introduction}

%%%%%%%%%%%%%%%%%%%%%%%%%%%%%%%%%%%%%%%%%%%%%%%%%%%%%%%%%%%%%%%%%%%

The Kochen-Specker (KS) theorem \cite{Specker60,KS67} shows that, for any quantum system of dimension $3$ or higher, the predictions of quantum theory (QT) cannot be reproduced with any theory that assumes the measurement results to be predefined and independent of other compatible measurements, i.e., noncontextual hidden variable (NCHV) theories, \cite{Peres95}. Bell's theorem \cite{Bell64} shows that, for entangled quantum states, the predictions of QT violate Bell inequalities satisfied by any theory that assumes the results of local measurements to be independent of measurements on spatially separated parts.

The proofs of both theorems are different. In the case of the KS theorem, the original proof consisted in a set of quantum yes-no tests, represented by rank-1 projectors, for which yes or no results cannot be assigned satisfying that, for every set of jointly measurable projectors, one and only one of the projectors can have assigned the result yes. The proof works for any quantum state of the system. In the case of Bell's theorem, the proof requires composite systems prepared in an entangled state and consists on the violation of a Bell inequality.

However, Kernaghan and Peres noticed that, for eight-dimensional quantum systems, there is a KS set of 40~yes-no quantum tests \cite{KP95} that is related \cite{Mermin90b,Mermin93} to Greenberger-Horne-Zeilinger's (GHZ) proof of Bell's theorem without inequalities \cite{GHZ89}, which can be reformulated as a violation of Mermin's Bell inequality \cite{Mermin90a} (which has been experimentally tested in, e.g., Refs.~\cite{PBDWZ00,EMFLHYPBPSRGLWJR14}).

Every Bell inequality can be converted into a noncontextuality (NC) inequality (i.e., one satisfied by any NCHV theory) involving sequential measurements (of compatible observables) on a single system rather than spacelike separated measurements (of compatible observables) on a composite system, and preserving both the compatibility relations existing in the Bell scenario and the maximum quantum violation. To see this, recall that two observables $A$ and $B$ are compatible if there exists an observable $M_{A,B}$ whose outcome set is the Cartesian product of the outcome sets of $A$ and $B$ and such that, for all states, the outcome probability distributions of $A$ and $B$ are recovered as marginals of the outcome probability distribution of $M_{A,B}$. If some observables are compatible then there exists a joint probability distribution for them. In scenarios where $A$ and $B$ are measured on separated systems, as in Bell inequality scenarios, constructing $M_{A,B}$ is immediate once one has local devices for measuring $A$ and $B$. In scenarios where arbitrary $A$ and $B$ are measured on the same system the problem is not so simple \cite{GKCLKZRR10}. However, if $A$ and $B$ are sharp quantum observables (i.e., quantum observables in von Neumann's sense \cite{VonNeumann32}), QT provides a prescription to build a measurement device for each of them \cite{RZBB94}. Then, a device for $M_{A,B}$ is simply one consisting of the devices for $A$ and $B$ placed sequentially in any order \cite{LKGC11}.

The aim of this article is to observe that, for different states, the violations of a NC inequality derived from Mermin's Bell inequality coincide with those predicted by QT for an experiment with spacelike measurements. Then, we experimentally show that this experiment is connected with Kernaghan and Peres's KS set.

For that, we start by experimentally implementing Kernaghan and Peres's KS set of tests using eight-dimensional quantum systems encoded in the transverse momentum of single photons, and observe the state-independent quantum violation of a noncontextuality inequality associated to the 40~KS tests. Then, we show that, if the system is prepared in a quantum state that is a single-system version of a GHZ state or in a state that is a single-system version of a $W$ state, the results of a subset of 16~KS tests violate a NC inequality which is formally equivalent to Mermin's Bell inequality, but for single eight-dimensional quantum systems. Moreover, for both states, the experimental violations of the NC inequality are in agreement with those predicted by QT for experiments with spacelike separated measurements.

In the sense explained before, our results provide {\em experimental} evidence of the connection between Bell and KS theorems and support the conclusion that quantum nonlocality, and its limits, are actually given by quantum contextuality and its limits. This may pave the way towards a deeper understanding of QT.

%%%%%%%%%%%%%%%%%%%%%%%%%%%%%%%%%%%%%%%%%%%%%%%%%%%%%%%%%%%%%%%%%%%

\section{The Kernaghan-Peres KS set}

%%%%%%%%%%%%%%%%%%%%%%%%%%%%%%%%%%%%%%%%%%%%%%%%%%%%%%%%%%%%%%%%%%%

\begin{figure*}[t]
\centering
\includegraphics[width=0.96 \textwidth]{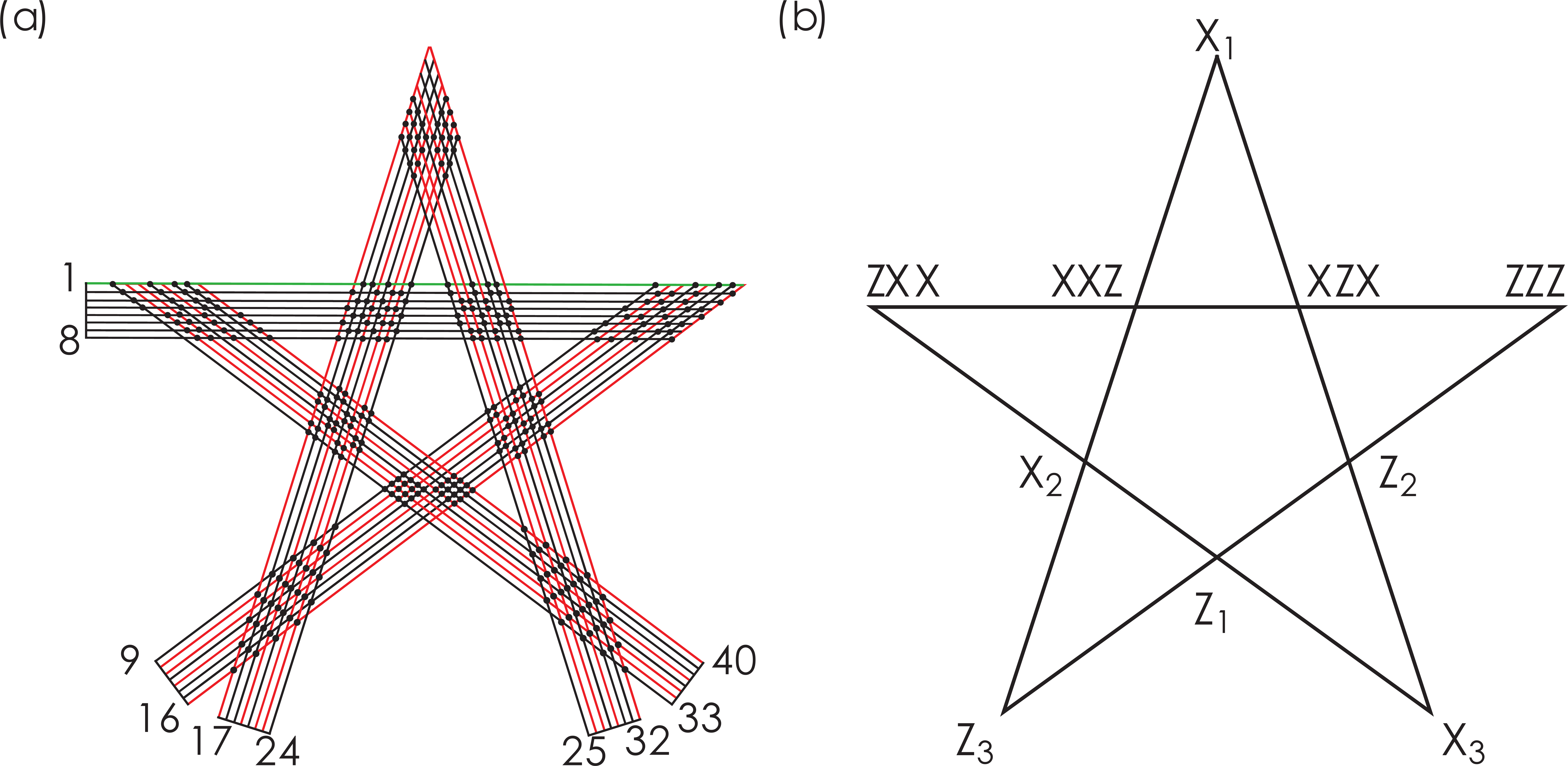}
\caption{(Color online) (a) The Kernaghan and Peres KS set of 40~quantum yes-no tests. Each test is represented by a straight line. The results of two tests cannot both be 1 when the lines are parallel or when there is a dot in their intersection. The initial state and the 16~KS tests violating the Bell inequality are indicated by a green line (the upper horizontal line) and 16~red (gray) lines, respectively. (b) Mermin's proof of state-independent quantum contextuality. There are ten observables represented as nodes in a pentagram. $ZXX$ denotes observable $\sigma_z \otimes \sigma_x \otimes \sigma_x$ on a system of three qubits, where $\sigma_z$ represents the Pauli $z$ matrix. Compatible observables are in the same line. Each of the 40~KS tests in (a) is the projection on a common eigenstate of 4 of the observables in (b). For example, states 1--8 in (a) are the common eigenstates of the observables in the horizontal line in (b) and similarly for the other lines. The impossibility of assigning non-contextual results $-1$ or $1$ to the observables in (b) in agreement with the predictions of QT leads to the impossibility of assigning non-contextual results~$1$ or $0$ to the yes-no tests in (a). \label{fig1}}
\end{figure*}

The KS set introduced by Kernaghan and Peres \cite{KP95} has 40~eight-dimensional vectors. Two yes-no tests that cannot both give result~$1$ (corresponding to ``yes''; result~$0$ corresponds to ``no'') are represented by orthogonal vectors. Table \ref{table1} shows these 40~vectors. The relations of orthogonality between the 40~vectors are represented in Fig.~\ref{fig1}(a).

%%%%%%%%%%%%%%%%%%%%%%%%%%%%%%%%%%%%%%%%%%%%%%%%%%%%%%%%%%%%%%%%%%%
% Table 1
%%%%%%%%%%%%%%%%%%%%%%%%%%%%%%%%%%%%%%%%%%%%%%%%%%%%%%%%%%%%%%%%%%%

\begin{table*}[tb]
\caption{Vectors representing the Kernaghan and Peres KS set of yes-no tests. The 40~vectors are shown in groups of eight elements which correspond to the eigenvectors of the commuting operators in Mermin's proof of quantum state-independent contextuality \cite{Mermin90b,Mermin93}. $x_1$ denotes $\sigma_x^{(1)}\otimes I \otimes I$, where $\sigma_x^{(1)}$ is the Pauli $x$ matrix for qubit 1 and $I$ is the two-dimensional identity matrix. $xzx$ denotes $\sigma_x^{(1)}\otimes\sigma_z^{(2)}\otimes\sigma_x^{(3)}$.}
\label{table1}
\begin{center}
\begin{tabular}{cclcclcclcclcc}
\hline \hline
\multicolumn{ 2}{c}{$\mathbf{zxx, xxz, xzx, zzz}$}
& & \multicolumn{ 2}{c}{$\mathbf{z_1,z_2,z_3,zzz}$}
& & \multicolumn{ 2}{c}{$\mathbf{x_1,z_2,z_3,xxz}$}
& & \multicolumn{ 2}{c}{$\mathbf{x_1,z_2,x_3,xzx}$}
& & \multicolumn{ 2}{c}{$\mathbf{x_1,x_2,x_3,zxx}$} \\
\hline
    \textbf{} & \textbf{}
& & \textbf{} & \textbf{}
& & \textbf{} & \textbf{}
& & \textbf{} & \textbf{}
& & \textbf{} & \textbf{} \\
\textbf{1:} & $(0,1,1,0,1,0,0,-1)$ & & \textbf{9:}  & $(1,0,0,0,0,0,0,0)$ & & \textbf{17:} & $(1,0,1,0,1,0,1,0)$   & & \textbf{25:} & $(0,0,1,-1,0,0,-1,1)$ & & \textbf{33:} & $(0,0,0,0,1,-1,-1,1)$ \\
\textbf{2:} & $(1,0,0,1,0,1,-1,0)$ & & \textbf{10:} & $(0,1,0,0,0,0,0,0)$ & & \textbf{18:} & $(0,1,0,1,0,1,0,1)$   & & \textbf{26:} & $(0,0,1,1,0,0,-1,-1)$ & & \textbf{34:} & $(0,0,0,0,1,1,-1,-1)$ \\
\textbf{3:} & $(1,0,0,1,0,-1,1,0)$ & & \textbf{11:} & $(0,0,1,0,0,0,0,0)$ & & \textbf{19:} & $(1,0,-1,0,1,0,-1,0)$ & & \textbf{27:} & $(1,-1,0,0,-1,1,0,0)$ & & \textbf{35:} & $(0,0,0,0,1,-1,1,-1)$ \\
\textbf{4:} & $(0,1,1,0,-1,0,0,1)$ & & \textbf{12:} & $(0,0,0,1,0,0,0,0)$ & & \textbf{20:} & $(0,1,0,-1,0,1,0,-1)$ & & \textbf{28:} & $(1,1,0,0,-1,-1,0,0)$ & & \textbf{36:} & $(0,0,0,0,1,1,1,1)$ \\
\textbf{5:} & $(1,0,0,-1,0,1,1,0)$ & & \textbf{13:} & $(0,0,0,0,1,0,0,0)$ & & \textbf{21:} & $(1,0,1,0,-1,0,-1,0)$ & & \textbf{29:} & $(0,0,1,-1,0,0,1,-1)$ & & \textbf{37:} & $(1,-1-1,1,0,0,0,0)$ \\
\textbf{6:} & $(0,1,-1,0,1,0,0,1)$ & & \textbf{14:} & $(0,0,0,0,0,1,0,0)$ & & \textbf{22:} & $(0,1,0,1,0,-1,0,-1)$ & & \textbf{30:} & $(0,0,1,1,0,0,1,1)$   & & \textbf{38:} & $(1,1,-1,-1,0,0,0,0)$ \\
\textbf{7:} & $(0,-1,1,0,1,0,0,1)$ & & \textbf{15:} & $(0,0,0,0,0,0,1,0)$ & & \textbf{23:} & $(1,0,-1,0,-1,0,1,0)$ & & \textbf{31:} & $(1,-1,0,0,1,-1,0,0)$ & & \textbf{39:} & $(1,-1,1,-1,0,0,0,0)$ \\
\textbf{8:} & $(-1,0,0,1,0,1,1,0)$ & & \textbf{16:} & $(0,0,0,0,0,0,0,1)$ & & \textbf{24:} & $(0,1,0,-1,0,-1,0,1)$ & & \textbf{32:} & $(1,1,0,0,1,1,0,0)$   & & \textbf{40:} & $(1,1,1,1,0,0,0,0)$ \\
\hline \hline
\end{tabular}
\end{center}
\end{table*}

%%%%%%%%%%%%%%%%%%%%%%%%%%%%%%%%%%%%%%%%%%%%%%%%%%%%%%%%%%%%%%%%%%%
% Fig. 1
%%%%%%%%%%%%%%%%%%%%%%%%%%%%%%%%%%%%%%%%%%%%%%%%%%%%%%%%%%%%%%%%%%%

%%%%%%%%%%%%%%%%%%%%%%%%%%%%%%%%%%%%%%%%%%%%%%%%%%%%%%%%%%%%%%%%%%%%

\section{A NC inequality violated by any eight-dimensional quantum state}

%%%%%%%%%%%%%%%%%%%%%%%%%%%%%%%%%%%%%%%%%%%%%%%%%%%%%%%%%%%%%%%%%%%

To show that the 40~tests prove the KS theorem for any eight-dimensional quantum state, we consider the following inequality satisfied by any NCHV theory:
\begin{equation}
 \label{IneKS40}
 \Sigma= \sum_{i=1}^{40} P(\Pi_i=1) \stackrel{\mbox{\tiny{ NCHV}}}{\leq} 4,
 \end{equation}
where $P(\Pi_i=1)$ is the probability of obtaining result~1 when performing test $\Pi_i=|v_i\rangle \langle v_i|$, and $|v_i\rangle$ are the KS vectors explicitly given in Table~\ref{table1}. Inequality (\ref{IneKS40}) follows from the observation that in any theory assigning a non-contextual result~$1$ or $0$ to each of the 40~tests $\Pi_i$, the maximum number of results~1 that can be assigned satisfying the relations of orthogonality in Fig.~\ref{fig1}(a) is 4.

However, in QT, for any initial quantum state,
\begin{equation}
 \Sigma \stackrel{\mbox{\tiny{Q}}}{=} 5.
\end{equation}
Therefore, state-independent quantum contextuality can be experimentally observed by showing that different initial eight-dimensional quantum states violate inequality (\ref{IneKS40}).

%%%%%%%%%%%%%%%%%%%%%%%%%%%%%%%%%%%%%%%%%%%%%%%%%%%%%%%%%%%%%%%%%%%

\section{A NC inequality formally equivalent to Mermin's inequality}

%%%%%%%%%%%%%%%%%%%%%%%%%%%%%%%%%%%%%%%%%%%%%%%%%%%%%%%%%%%%%%%%%%%

For three-qubit systems, Mermin's Bell inequality \cite{Mermin90a} for local hidden variable (LHV) theories states that
\begin{equation}
\kappa= \langle zxx \rangle + \langle xzx \rangle + \langle xxz \rangle - \langle zzz \rangle \stackrel{\mbox{\tiny{ LHV}}}{\leq} 2, \label{Mermin_ini}
\end{equation}
where $\langle zxx \rangle$ is the mean value of the product of measuring observable $z$ (with possible results $-1$ or $+1$) on qubit $1$, $x$ on qubit $2$, and $x$ on qubit $3$. Inequality (\ref{Mermin_ini}) can be rewritten as
\begin{eqnarray}
 \label{Merm_Ineq}
S &=& \frac{\kappa}{2}+2 = P(z_1=1,x_2=1,x_3=1)+\ldots\nonumber \\
& &+P(z_1=-1,z_2=-1,z_3=-1) \stackrel{\mbox{\tiny{ NCHV}}}{\leq} 3,
\end{eqnarray}
where $P(z_1=1,x_2=1,x_3=1)$ is the probability of obtaining result~1 when $z$ is measured on qubit~$1$, $x$ is measured on qubit~$2$, and is measured $x$ on qubit~$3$. The 16~probabilities in $S$ are the probabilities of the 12~events in which the product of the results of $z_i$, $x_j$, and $x_k$, with $i \neq j \neq k \neq i$, is $1$ and the probabilities of the four events in which the product of the results of $z_1$, $z_2$, and $z_3$ is $-1$.

On the other hand, notice that a three-qubit system is an eight-dimensional quantum system. Therefore, one can see that $S$ is the sum of 16~probabilities that also appear in inequality (\ref{IneKS40}), and thus $S$ constitutes a new NC inequality that is the single-particle equivalent of Mermin's Bell inequality. Table~\ref{table2} shows the 16~yes-no tests appearing in this NC inequality. The reason why the NC Mermin inequality state projections coincide with the states of the Kernaghan-Peres KS set follows from three observations: first, that GHZ's proof of Bell's theorem \cite{GHZ89} can be converted into a violation of a Bell inequality \cite{Mermin90a}; second, that GHZ's proof can be extended into a proof of state-independent quantum contextuality \cite{Mermin90b,Mermin93} [this last proof is shown in Fig.~\ref{fig1}(b)]; and third, that this last proof is connected to a proof of the KS theorem \cite{KP95} (see Fig.~\ref{fig1}).

%%%%%%%%%%%%%%%%%%%%%%%%%%%%%%%%%%%%%%%%%%%%%%%%%%%%%%%%%%%%%%%%%%%
% Table 2
%%%%%%%%%%%%%%%%%%%%%%%%%%%%%%%%%%%%%%%%%%%%%%%%%%%%%%%%%%%%%%%%%%%

\begin{table*}[tb]
\caption{The 16 yes-no tests in the NC Mermin inequality.}
\label{table2}
\begin{center}
\begin{tabular}{cclcclcclcc}
\hline \hline
    \multicolumn{ 2}{c}{$\mathbf{z_1,z_2,z_3,zzz}$}
& & \multicolumn{ 2}{c}{$\mathbf{x_1,z_2,z_3,xxz}$}
& & \multicolumn{ 2}{c}{$\mathbf{x_1,z_2,x_3,xzx}$}
& & \multicolumn{ 2}{c}{$\mathbf{x_1,x_2,x_3,zxx}$} \\
\hline
    \textbf{} & \textbf{}
& & \textbf{} & \textbf{}
& & \textbf{} & \textbf{}
& & \textbf{} & \textbf{} \\
\textbf{10:} & $(0,1,0,0,0,0,0,0)$ & & \textbf{17:} & $(1,0,1,0,1,0,1,0)$ & & \textbf{26:} & $(0,0,1,1,0,0,-1,-1)$ & & \textbf{34:} & $(0,0,0,0,1,1,-1,-1)$ \\
\textbf{11:} & $(0,0,1,0,0,0,0,0)$ & & \textbf{20:} & $(0,1,0,-1,0,1,0,-1)$ & & \textbf{27:} & $(1,-1,0,0,-1,1,0,0)$ & & \textbf{35:} & $(0,0,0,0,1,-1,1,-1)$ \\
\textbf{13:} & $(0,0,0,0,1,0,0,0)$ & & \textbf{22:} & $(0,1,0,1,0,-1,0,-1)$ & & \textbf{29:} & $(0,0,1,-1,0,0,1,-1)$ & & \textbf{37:} & $(1,-1-1,1,0,0,0,0)$ \\
\textbf{16:} & $(0,0,0,0,0,0,0,1)$ & & \textbf{23:} & $(1,0,-1,0,-1,0,1,0)$ & & \textbf{32:} & $(1,1,0,0,1,1,0,0)$ & & \textbf{40:} & $(1,1,1,1,0,0,0,0)$ \\
\hline \hline
\end{tabular}
\end{center}
\end{table*}

%%%%%%%%%%%%%%%%%%%%%%%%%%%%%%%%%%%%%%%%%%%%%%%%%%%%%%%%%%%%%%%%%%%

Now, when we prepare an eight-dimensional system in one specific vector of the KS set, that is, an initial state for which the first KS test gives result~1 (the GHZ state), and perform the projections for the 16~KS tests in the NC Mermin inequality, we obtain that
\begin{equation}
S \stackrel{\mbox{\tiny{Q}}}{=} 4,
\end{equation}
maximally violating NC inequality (\ref{Merm_Ineq}). The observation of such a violation constitutes an experimental observation that there is a connection between quantum contextuality and nonlocality for systems of dimension~$8$. This is due to the aforementioned relation between inequalities (\ref{IneKS40}), (\ref{Mermin_ini}), and (\ref{Merm_Ineq}). Here it is important to clarify that the advantage of performing the tests of inequalities (\ref{IneKS40}) and (\ref{Merm_Ineq}) with tripartite entangled states is that a single experiment can refute LHV and NCHV theories, simultaneously. However, as we show in the next section, single eight-dimensional systems suffice for observing the connection between the KS and GHZ theorems.

%%%%%%%%%%%%%%%%%%%%%%%%%%%%%%%%%%%%%%%%%%%%%%%%%%%%%%%%%%%%%%%%%%%

\section{Experiment}

%%%%%%%%%%%%%%%%%%%%%%%%%%%%%%%%%%%%%%%%%%%%%%%%%%%%%%%%%%%%%%%%%%%

In our experiment, the eight-dimensional quantum states are encoded in the linear transverse momentum of single photons transmitted by diffractive apertures addressed in spatial light modulators (SLMs) \cite{Twi1,Moreno_opex}. The dimension of the quantum states is defined by the number of paths available for the photon transmission \cite{Neves05,Lima09,Lima11,Lima13}. In our case, we have an aperture composed of eight parallel slits, and the state of the transmitted photons is given by \cite{Neves05,Lima09}
\begin{equation}
\label{State}
\ket{\Psi} = \frac{1}{\sqrt{N}}\sum_{l=-\frac{7}{2}}^{\frac{7}{2}} \sqrt{t_{l}}e^{i\phi_l}\ket{l},
\end{equation}
where $\ket{l}$ represents the state of a photon transmitted by the $l$th-slit \cite{Neves05}. $t_l$ ($\phi_l$) is the transmissivity (phase) defined for each slit, and $N$ is the normalization constant.

%%%%%%%%%%%%%%%%%%%%%%%%%%%%%%%%%%%%%%%%%%%%%%%%%%%%%%%%%%%%%%%%%%%
% Fig. 2
%%%%%%%%%%%%%%%%%%%%%%%%%%%%%%%%%%%%%%%%%%%%%%%%%%%%%%%%%%%%%%%%%%%

\begin{figure}[t]
\includegraphics[width=0.48\textwidth]{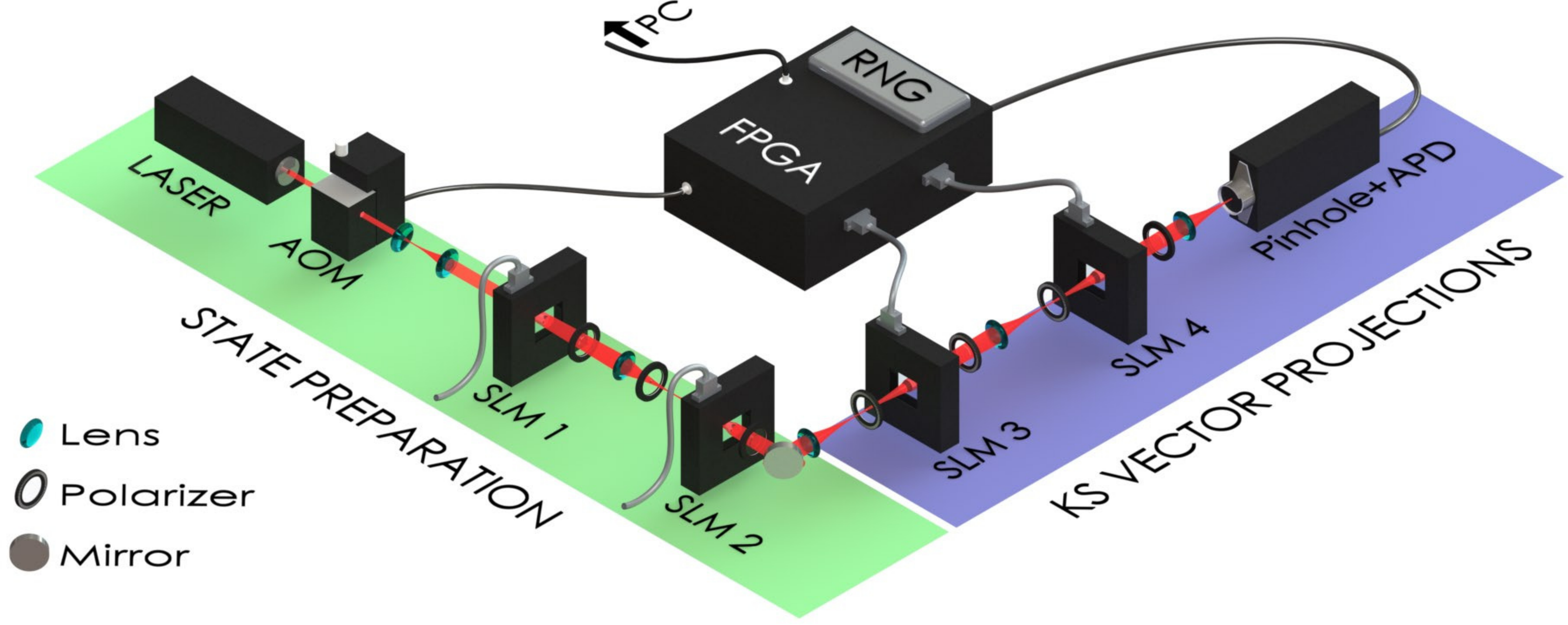}
\caption{(Color online) Experimental setup. A CW laser, an AOM and calibrated attenuators (not shown for clarity) are used to produce faint optical pulses. The weak coherent states are transversally expanded by a telescope and sent through four transmissive SLMs placed in series, and with each LCD at the image plane of the previous one. SLM1 and SLM2 are used to prepare initial eight-dimensional quantum states encoded in the linear transverse momentum of single photons. The generated states are then used to test inequalities (\ref{IneKS40}) and (\ref{Merm_Ineq}) after performing, on each of them, the 40~KS vector projections. The KS projections are carried out using SLM3, SLM4 and a point-like APD (see the main text for details).
\label{fig2}}
\end{figure}

%%%%%%%%%%%%%%%%%%%%%%%%%%%%%%%%%%%%%%%%%%%%%%%%%%%%%%%%%%%%%%%%%%%

The experimental setup is depicted in Fig.~\ref{fig2}. A single-mode continuous-wave (CW) laser operating at $690$ nm is combined with an acousto-optic modulator (AOM) to produce attenuated optical pulses with a mean photon number of $\mu = 0.14$. The first two SLMs, SLM1 and SLM2, are used to prepare the initial state, $\ket{\Psi_{\mathrm{ini}}}$. SLM1 (SLM2) displays an amplitude (phase) mask of eight slits with the gray level of the used pixels properly set for the generation of the desired initial state. The slits are 2 pixels wide, with a separation of 1 pixel between them, where each pixel is a square with a side length of $32$ $\mu$m. The projections onto the vectors of the KS set are carried out using a second pair of SLMs, SLM3 and SLM4, and a point-like avalanche photo-detector (APD). The masks of these last modulators have the same dimension as the ones used by SLM1 and SLM2. The only difference between them is that of the gray levels used, now set according to the amplitudes and phases of the 40~KS vectors. After the SLM4, the attenuated laser beam is focused at the detection plane. The point-like detector is constructed using a pinhole in front of a conventional bulk APD, which is then positioned at the center of the interference pattern. In this configuration, the probability of single-photon detection at the APD is proportional to $|\bra{v_i}\Psi_{\mathrm{ini}} \rangle |^2$ \cite{Lima11,Lima13,Leo10,Lima10}.

During the implementation of the KS tests, the modulations performed by SLM1 and SLM2 remain fixed producing the quantum state to be used in the test. The AOM, SLM3, SLM4, and APD are connected to a field programmable gate array (FPGA) electronics unit, which synchronizes the optical pulse generation, the masks displayed in both SLMs, and the detection time of the APD. This synchronization allows us to project, for each optical pulse, the prepared state into a different KS vector. The projection is randomly chosen by using a true random number generator (RNG) connected to the FPGA. For each initial state considered, the setup automatically runs for 17~h (with more than $2 \times 10^6$ detected pulses) in order to minimize statistical fluctuations and unambiguously certify its quantum behavior (see Appendix~\ref{AppendixA}).

\begin{figure}[t]
\centering
\includegraphics[width=0.36\textwidth]{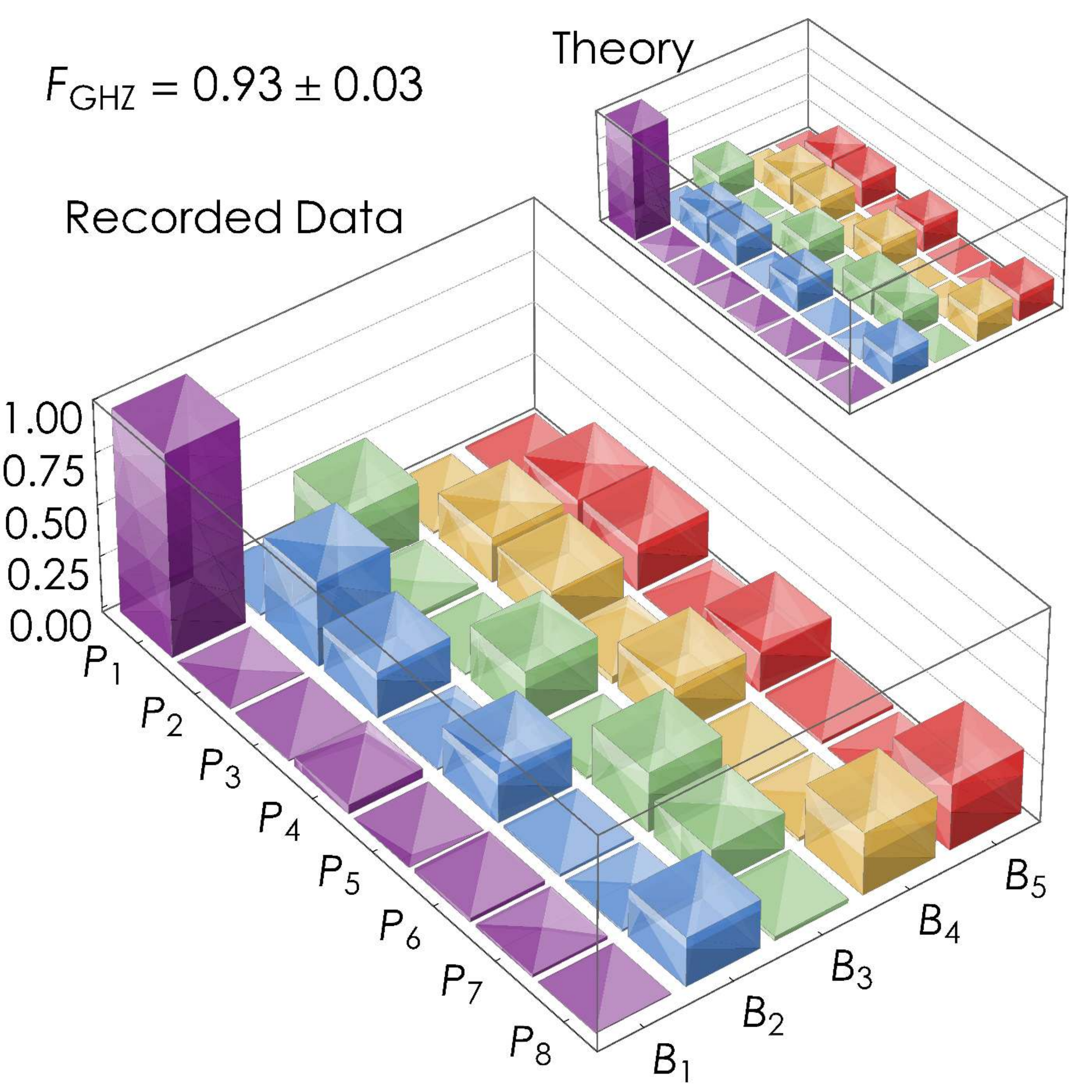}
\caption{(Color online) Probabilities of result~1 for the 40~KS tests in $\Sigma$ when the initial state is the GHZ state. The 40~KS tests are grouped in five bases, as in Fig.~\ref{fig1}(a). The theoretical predictions for an ideal experiment are shown in the upper right corner. $F$ is the similarity of the recorded and theoretical probability distributions.
\label{fig3}}
\end{figure}

State-independent quantum contextuality is observed through the violation of inequality (\ref{IneKS40}) for five different types of initial quantum states. Namely, a GHZ state [$\langle \mathrm{GHZ} | \equiv \frac{1}{2} (0, 1, 1, 0, 1, 0, 0, -1)$] \cite{GHZ89}, a $W$ state [$\langle W | \equiv \frac{1}{\sqrt{3}}(0,1,1,0,1,0,0,0)$] \cite{DVC00}, a state
that is equivalent to a product of a two-qubit maximally entangled state and a pure state of one qubit [$\langle \beta | \equiv \frac{1}{2}(0,0,1,1,-1,-1,0,0)$], a state equivalent to a product of a two-qubit partially entangled state and a pure state of one qubit [$\langle \eta | \equiv \frac{1}{\sqrt{6}}(1,1,1,1,1,1,0,0)$], and a product state [$\langle \Psi_{\mathrm{prod}} | \equiv (1,0,0,0,0,0,0,0)$]. In Fig.~\ref{fig3} we compare the theoretical and the experimental results for all the 40~projection probabilities appearing in $\Sigma$ when the initial state of the system is the GHZ state. The similarity, $F$, of the recorded and theoretical probability distributions reaches $F_{\mathrm{GHZ}}=0.93 \pm 0.03$ \cite{Bhatta} (see Appendix~\ref{AppendixB} for details of the other four initial states).

%%%%%%%%%%%%%%%%%%%%%%%%%%%%%%%%%%%%%%%%%%%%%%%%%%%%%%%%%%%%%%%%%%%
% Fig. 3
%%%%%%%%%%%%%%%%%%%%%%%%%%%%%%%%%%%%%%%%%%%%%%%%%%%%%%%%%%%%%%%%%%%

%%%%%%%%%%%%%%%%%%%%%%%%%%%%%%%%%%%%%%%%%%%%%%%%%%%%%%%%%%%%%%%%%%%

\begin{figure}[t]
\centering
\includegraphics[width=0.50\textwidth]{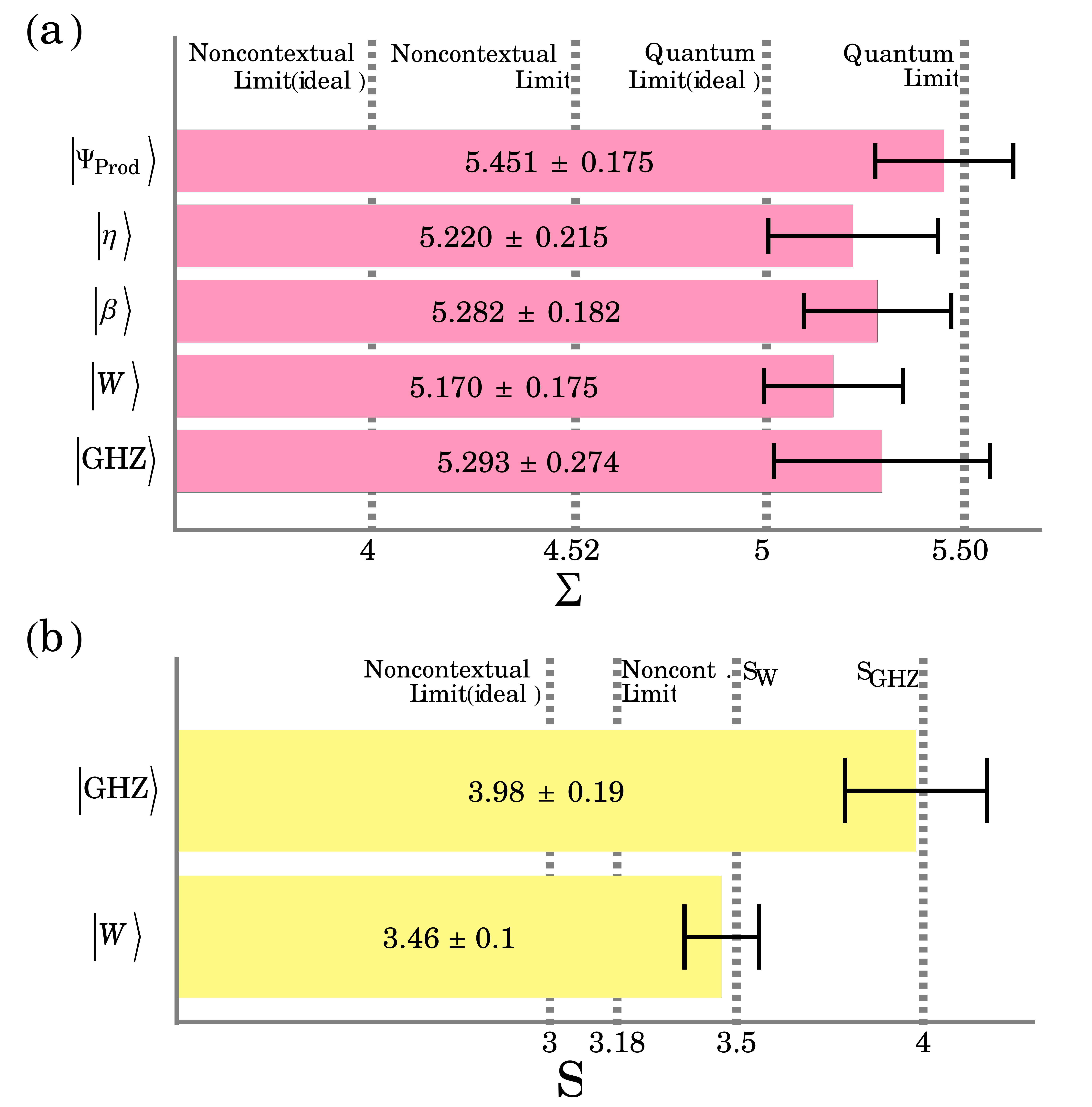}
\caption{(Color online) (a) Experimental results for $\Sigma$. The dotted line ``Noncontextual limit (ideal)'' indicates the maximum possible value of $\Sigma$ for noncontextual hidden variable theories in an ideal experiment in which the exclusivity relations between the KS tests are perfect. The dotted line ``Noncontextual limit'' indicates the value when experimental imperfections are taken into account, and similarly for the quantum limits. (b) Experimental results for $S$. The measured violations for the GHZ and $W$ states are within the theoretical predictions for an ideal experiment, namely, $S_{\mathrm{GHZ}}=4.0$ and $S_{W}=3.5$, respectively.
\label{fig4}}
\end{figure}

Due to intrinsic experimental imperfections in real experiments, the mean values of the recorded probabilities that were supposed to be null, in an ideal experiment, do not vanish. Thus, it is necessary to modify the noncontextual limit of inequality (\ref{IneKS40}) to proper demonstrate quantum contextuality. In this work we follow the approach of \cite{4dimKStest}. The idea is simple: first one takes from the experimental data the mean value of the recorded probabilities that were supposed to be null (i.e., $\epsilon$). To measure it, one needs to perform the so-called exclusivity tests, where one of the KS vectors is used as the initial state and the remaining orthogonal measurements are performed. To properly estimate the value of $\epsilon$, we have performed a total of 184 exclusivity tests and obtained that $\epsilon=0.0140 \pm 0.0012$ (see Appendix~\ref{AppendixC} for some examples of the exclusivity tests performed). Then, since the inequality \cite{4dimKStest} involves a sum of all the KS vector projection probabilities, one has a noncontextual upper bound given by $\Sigma_{\mathrm{upper}}^{\mathrm{clas}}=4(1-\epsilon)+40\epsilon$, where the noncontextual limit of $4$ is achieved with probability $1-\epsilon$, and all the 40~KS tests give false positive results with probability $\epsilon$. In our case, $\Sigma_{\mathrm{upper}}^{\mathrm{clas}} = 4.52$. The results obtained for the five initial states considered are shown in Fig.~\ref{fig4}(a). All of them clearly violate the noncontextual limit, demonstrating the impossibility of non-contextual hidden variables models explaining our results. The quantum limit when errors are taken into account is similarly obtained \cite{4dimKStest}.

%%%%%%%%%%%%%%%%%%%%%%%%%%%%%%%%%%%%%%%%%%%%%%%%%%%%%%%%%%%%%%%%%%%
% Fig. 4
%%%%%%%%%%%%%%%%%%%%%%%%%%%%%%%%%%%%%%%%%%%%%%%%%%%%%%%%%%%%%%%%%%%

%%%%%%%%%%%%%%%%%%%%%%%%%%%%%%%%%%%%%%%%%%%%%%%%%%%%%%%%%%%%%%%%%%%

For the GHZ \cite{GHZ89} and the $W$ \cite{DVC00} states we also record the corresponding violation of the noncontextual limit of the NC Mermin inequality. The obtained violations are shown in Fig.~\ref{fig4}(b). For the GHZ state the measured value was $S=3.98\pm0.19$. For the $\ket{W}$ state, the violation was $S=3.46 \pm 0.1$. In the case of the GHZ state, the observed violation demonstrates that, if the answer to one of the KS tests is positive, no noncontextual hidden variable model can explain the obtained results of the 16~KS tests in the NC Mermin inequality. Notice that our experiment is a test of a NC inequality formally equivalent to a Bell inequality in the sense of previous tests on single systems \cite{Hasegawa03}. It is for this reason that the noncontextual limit of inequality (\ref{Merm_Ineq}) must be modified to take errors into account, using the same argument used to correct the noncontextual limit of inequality (\ref{IneKS40}). In our case, the experimental value, $\epsilon=0.0140$, modifies the noncontextual limit to $3.18$. Fig.~\ref{fig4}(b) shows that the experimental results violate this limit.

%%%%%%%%%%%%%%%%%%%%%%%%%%%%%%%%%%%%%%%%%%%%%%%%%%%%%%%%%%%%%%%%%%%

\section{Conclusions}

%%%%%%%%%%%%%%%%%%%%%%%%%%%%%%%%%%%%%%%%%%%%%%%%%%%%%%%%%%%%%%%%%%%

Recent experiments have shown that KS sets of quantum tests can be used to experimentally reveal quantum state-independent contextuality for quantum systems of a given dimension. Previous experiments have shown this for quantum systems of dimension 3 \cite{Zhang13} and 4 \cite{4dimKStest}. Here we have implemented a KS set in dimension 8 and shown how to use it to reveal eight-dimensional quantum state-independent contextuality through the violation of a NC inequality.

The KS set we have implemented is particularly important because it connects KS and GHZ proofs of no hidden variables. Specifically, GHZ's proof can be seen as a state-dependent version of the proof of the KS theorem. Here we have experimentally shown this connection by preparing eight-dimensional single systems in a state that is formally equivalent to a GHZ state, and observing a violation of a NC inequality that is a single-system version of Mermin's inequality. Our results also show how highly sophisticated theoretical tools can be translated into actual experiments to test fundamental aspects of QT.

%%%%%%%%%%%%%%%%%%%%%%%%%%%%%%%%%%%%%%%%%%%%%%%%%%%%%%%%%%%%%%%%%%%%

\section*{Acknowledgments}

%%%%%%%%%%%%%%%%%%%%%%%%%%%%%%%%%%%%%%%%%%%%%%%%%%%%%%%%%%%%%%%%%%%%

This work was supported by Grant No.\ FONDECYT 1120067, No.\ CONICYT PFB08-024, and No.\ Milenio P10-030-F (Chile); Project No.\ FIS2011-29400 (MINECO, Spain) with FEDER funds; the FQXi large grant project ``The Nature of Information in Sequential Quantum Measurements''; and the Brazilian program Science without Borders. G.\ B.\ X.\ acknowledges financial support from Grant No.\ FONDECYT 11110115. G.\ C.\ and E.\ S.\ G.\ also acknowledge the financial support of CONICYT and AGCI.

%%%%%%%%%%%%%%%%%%%%%%%%%%%%%%%%%%%%%%%%%%%%%%%%%%%%%%%%%%%%%%%%%%%%

\appendix

%%%%%%%%%%%%%%%%%%%%%%%%%%%%%%%%%%%%%%%%%%%%%%%%%%%%%%%%%%%%%%%%%%%%

\section{Measurement process and results}
\label{AppendixA}

%%%%%%%%%%%%%%%%%%%%%%%%%%%%%%%%%%%%%%%%%%%%%%%%%%%%%%%%%%%%%%%%%%%%

\begin{figure*}[tb]
\centering
\includegraphics[width=0.85\textwidth]{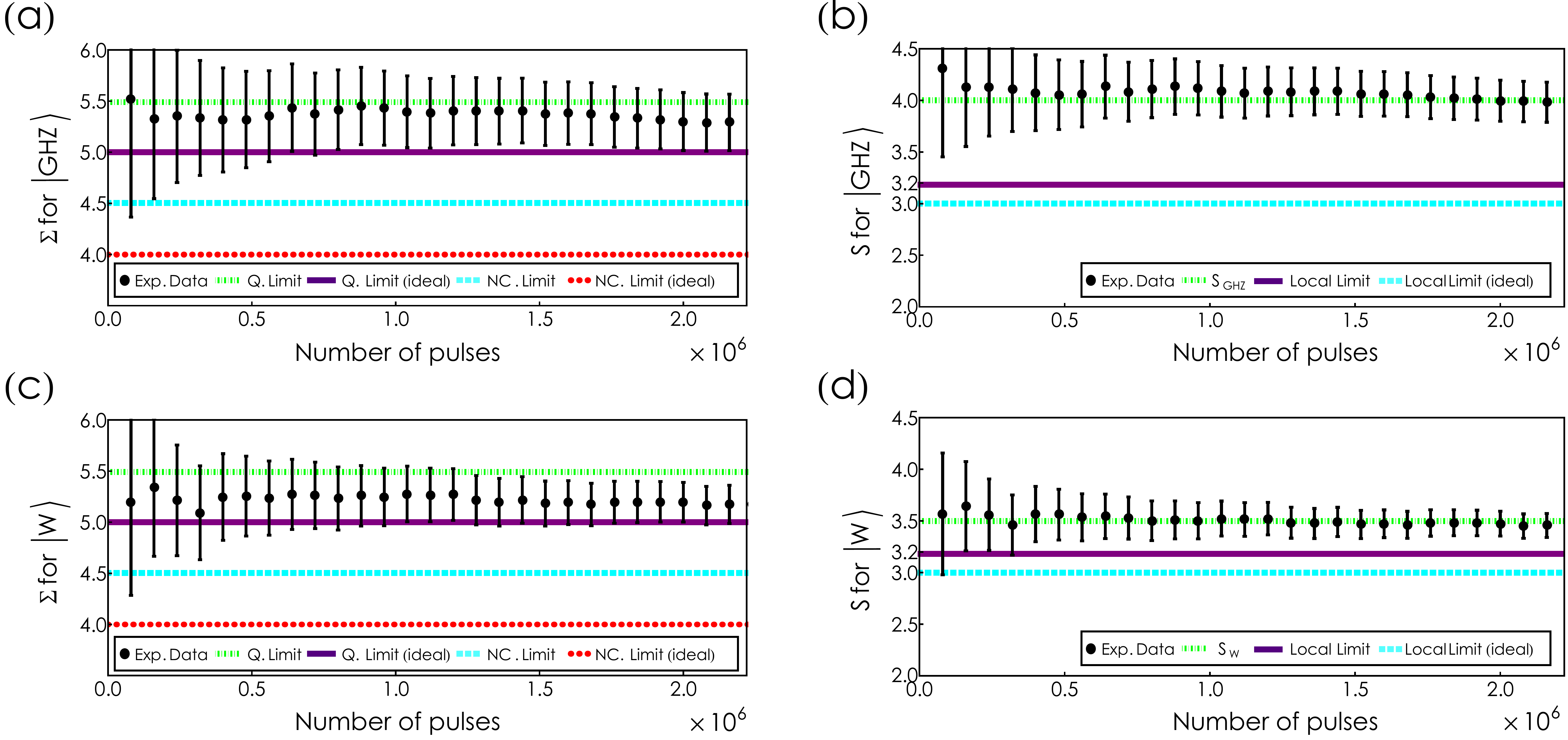}
\caption{(Color online) Figure (a) [(c)] shows the measured value of $\Sigma$ for a certain number of detected pulses for the $|\mathrm{GHZ}\rangle$ ($|W\rangle$) state. Figure (b) [(d)] shows the measured value of the NC Mermin inequality violation for a certain number of detected pulses for the $|\mathrm{GHZ}\rangle$ ($|W\rangle$) state.}\label{Fig1}
\end{figure*}

\begin{figure*}[tb]
\centering
\includegraphics[width=0.85\textwidth]{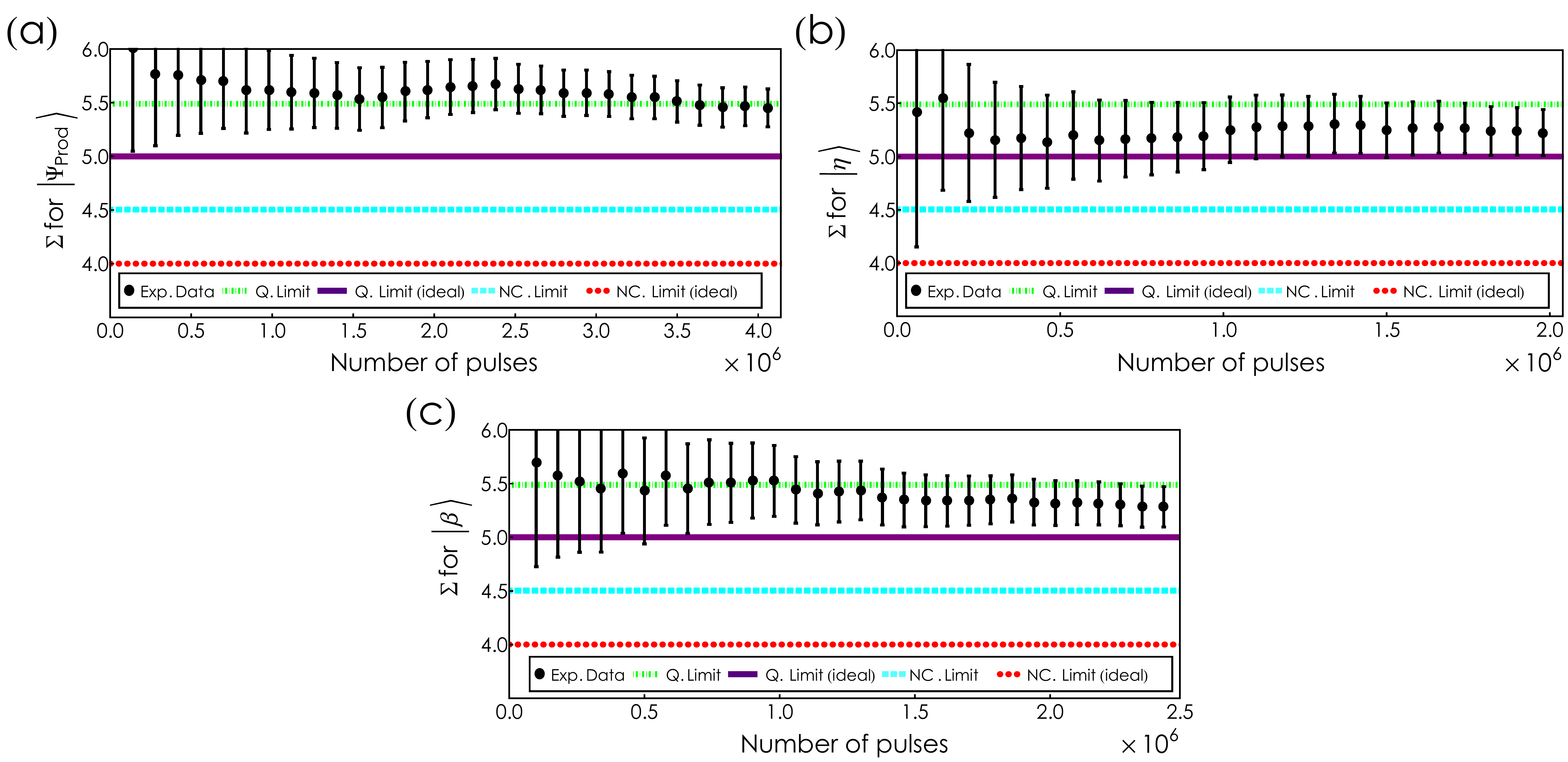}
\caption{(Color online) Figure (a), (b), and (c) show the measured value of $\Sigma$ for a certain number of detected pulses for the $|\Psi_{\mathrm{prod}}\rangle$, $|\eta\rangle$, and $|\beta\rangle$ states, respectively.}\label{Fig2}
\end{figure*}

The measurements performed for the five different initial states, namely, $|\mathrm{GHZ}\rangle$, $|W\rangle$, $|\Psi_{\mathrm{prod}}\rangle$, $|\eta\rangle$ and $|\beta\rangle$, are shown in Figs.~\ref{Fig1} and \ref{Fig2}. As explained in the main text, the experimental setup projects the initial state randomly onto one of the 40~KS vector states of Table~\ref{table1}. For each optical pulse, a random projection is implemented using a field programmable gate array (FPGA) electronics unit. The FPGA synchronizes the optical pulse generation, the masks displayed in SLM3 and SLM4, and the detection time of the APD. The detected counts are sent from the FPGA to a computer for real time estimation of the probabilities of each projection and the corresponding errors. The error bars were calculated taking into account the Poissonian distribution for the single counts recorded. The experimental setup automatically runs for 17~h (with more than $2 \times 10^6$ detected pulses), for each initial state, in order to minimize statistical fluctuations and unambiguously certify the quantum behavior of the considered states. The results shown in Fig.~\ref{fig4} correspond to the last experimental points of Figs.~\ref{Fig1} and \ref{Fig2}.

%%%%%%%%%%%%%%%%%%%%%%%%%%%%%%%%%%%%%%%%%%%%%%%%%%%%%%%%%%%%%%%%%%%%
% Fig. 5
%%%%%%%%%%%%%%%%%%%%%%%%%%%%%%%%%%%%%%%%%%%%%%%%%%%%%%%%%%%%%%%%%%%%

%%%%%%%%%%%%%%%%%%%%%%%%%%%%%%%%%%%%%%%%%%%%%%%%%%%%%%%%%%%%%%%%%%%%
% Fig. 6
%%%%%%%%%%%%%%%%%%%%%%%%%%%%%%%%%%%%%%%%%%%%%%%%%%%%%%%%%%%%%%%%%%%%

%%%%%%%%%%%%%%%%%%%%%%%%%%%%%%%%%%%%%%%%%%%%%%%%%%%%%%%%%%%%%%%%%%%%

\section{Probabilities for the KS tests}
\label{AppendixB}

%%%%%%%%%%%%%%%%%%%%%%%%%%%%%%%%%%%%%%%%%%%%%%%%%%%%%%%%%%%%%%%%%%%%

\begin{figure*}[t]
\centering
\includegraphics[width=0.7\textwidth]{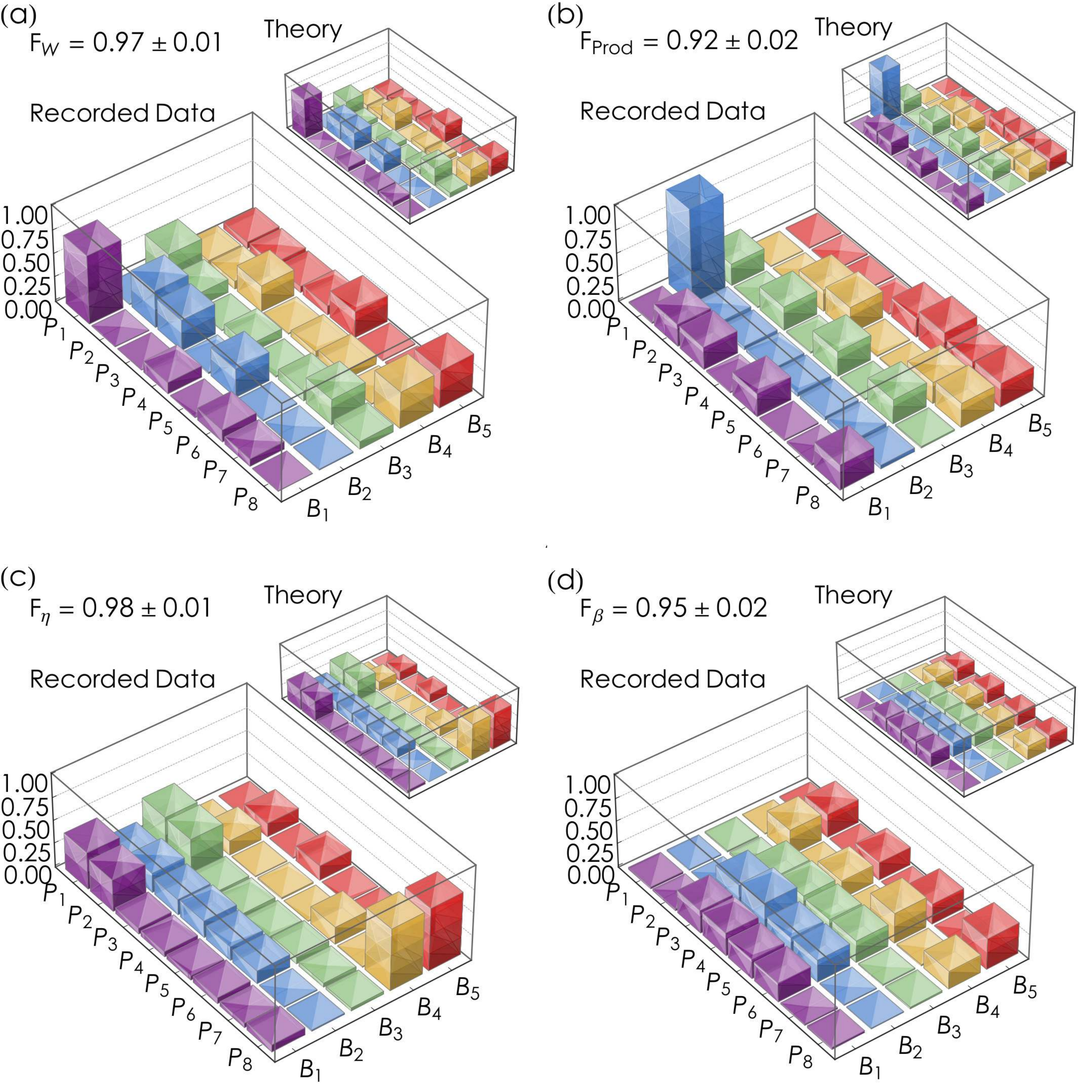}
\caption{(Color online) Figure (a), (b), (c), and (d) show the probabilities of result~1 for the 40~KS tests performed for the $|W\rangle$, $|\Psi_{\mathrm{prod}}\rangle$, $|\eta\rangle$ and $|\beta\rangle$ states, respectively. The fidelities, $F$, between the recorded and the expected probability distributions are $0.97 \pm 0.01$, $0.92 \pm 0.02$, $0.98 \pm 0.01$, and $0.95 \pm 0.02$, respectively.\label{Fig3}}
\end{figure*}

In Fig.~\ref{Fig3} we compare the theory and the experimental results for the probabilities of obtaining result~1 for the 40~KS tests, while considering the following initial states: $|W\rangle$, $|\Psi_{\mathrm{prod}}\rangle$, $|\eta\rangle$, and $|\beta\rangle$. The probabilities are obtained by normalizing the corresponding counts with respect to the total number of photons being detected per base, while working with a fixed detection time. Note, however, that the total number of photons being registered in an overcomplete basis must be estimated independently of the KS tests (in our case, by using additional measurement configurations), such that the violation of the noncontextual inequality for $\Sigma$ is not intrinsically imposed.

%%%%%%%%%%%%%%%%%%%%%%%%%%%%%%%%%%%%%%%%%%%%%%%%%%%%%%%%%%%%%%%%%%%%
% Fig. 7
%%%%%%%%%%%%%%%%%%%%%%%%%%%%%%%%%%%%%%%%%%%%%%%%%%%%%%%%%%%%%%%%%%%%

%%%%%%%%%%%%%%%%%%%%%%%%%%%%%%%%%%%%%%%%%%%%%%%%%%%%%%%%%%%%%%%%%%%%

\section{Tests of exclusivity}
\label{AppendixC}

%%%%%%%%%%%%%%%%%%%%%%%%%%%%%%%%%%%%%%%%%%%%%%%%%%%%%%%%%%%%%%%%%%%%

\begin{figure*}[tb]
\centering
\includegraphics[width=0.8\textwidth]{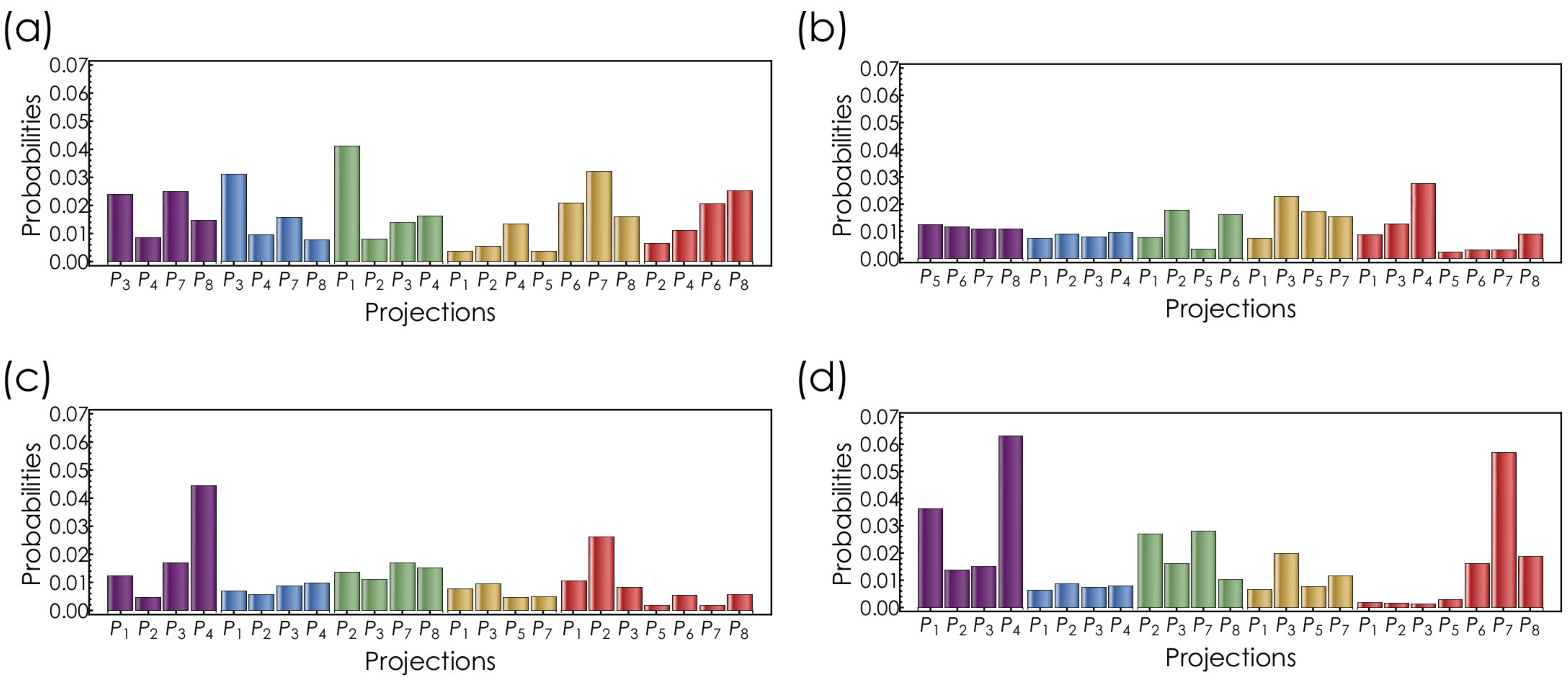}
\caption{(Color online) Figure (a), (b), (c), and (d) show the projection probabilities for the KS vectors orthogonal to the $|\Psi_\mathrm{v_{27}}\rangle$, $|\Psi_\mathrm{v_{34}}\rangle$, $|\Psi_\mathrm{v_{36}}\rangle$, and $|\Psi_\mathrm{v_{40}}\rangle$ states, respectively. The mean values of the probabilities shown are $0.016\pm 0.002, 0.011\pm 0.001, 0.011\pm 0.001$, and $0.017\pm 0.001$, respectively. \label{Fig4}}
\end{figure*}

Some pairs of yes-no tests in the KS set have to satisfy that their results cannot be simultaneously~1. In Fig.~\ref{Fig4} we show the results of some of the tests performed to check whether the KS yes-no tests actually satisfy the expected relations of exclusivity. The mean value of the recorded probabilities that were supposed to be null is $\epsilon=0.0140\pm 0.0012$. This is the parameter used to correct the noncontextual upper bounds, as explained in the main text. For testing exclusivity, one of the KS vectors is used as the initial state and the projection probabilities on each of the 23 KS vectors that are expected to be orthogonal are recorded.

%%%%%%%%%%%%%%%%%%%%%%%%%%%%%%%%%%%%%%%%%%%%%%%%%%%%%%%%%%%%%%%%%%%%
% Fig. 8
%%%%%%%%%%%%%%%%%%%%%%%%%%%%%%%%%%%%%%%%%%%%%%%%%%%%%%%%%%%%%%%%%%%%

%%%%%%%%%%%%%%%%%%%%%%%%%%%%%%%%%%%%%%%%%%%%%%%%%%%%%%%%%%%%%%%%%%%

%%%%%%%%%%%%%%%%%%%%%%%%%%%%%%%%%%%%%%%%%%%%%%%%%%%%%%%%%%%%%%%%%%%

\end{document}